\documentclass{eas}
\usepackage{graphicx}
\TitreGlobal{Physics and Astrophysics of Planetary Systems, Les Houches 2008}
\begin{document}

\title{Planetary migration in resonance\,:\\ the question of the eccentricities} 
\runningtitle{Crida \etal: Migration in resonance}
\author{Aur\'elien Crida}\address{Institut f\"ur Astronomie \& Astrophysik -- Universit\"at T\'ubingen -- Auf der Morgenstelle 10 -- 72076 {\sc T\"ubingen} -- GERMANY}
\author{Zsolt S\'andor}\address{Max-Planck-Institut f\"ur Astronomie -- K\"onigstuhl 17 -- 69711 {\sc Heidelberg} -- GERMANY}
\author{Wilhelm Kley}\sameaddress{1}
%

%
\begin{abstract}
The formation of resonant planets pairs in exoplanetary systems
involves planetary migration inside the protoplanetary disc\,: an
inwards migrating outer planet captures in Mean Motion Resonance an
inner planet. During the migration of the resonant pair of planets,
the eccentricities are expected to rise excessively, if no damping
mechanism is applied on the inner planet. We express the required
damping action to match the observations, and we show that the inner
disk can play this role. This result applies for instance to the
system GJ~876\,: we reproduce the observed orbital elements through a
fully hydrodynamical simulation of the evolution of the resonant
planets.
\end{abstract}
\maketitle
\section{Introduction}
Most of the known exoplanets are giant planets on short period orbits
around their stars. In the core-accretion model, such planets should
have formed much further away, beyond the so-called ``snow-line'',
where water is solid and contributes to the formation of a massive
solid core. This suggests that inward planetary migration occured in
the protoplanetary disc. Through mutual gravitationnal interactions, a
planet and the gaseous disc exchange energy and angular momentum. For
a giant planet, this lead to the opening of a gap and to the migration
of the planet (\cite{LinPapa86}).

A stronger evidence for migration is that a large fraction of
multiplanet systems are engaged a Mean Motion Resonance (MMR)\,: the
ratio of the orbital periods of the outer to the inner planet,
equals that of two small integers. This cannot happen by chance. It
betrays a convergent migration of the planets and a dissipative
process to enter this dynamical configuration.

For instance, in GJ~876, the two most massive planets are in a 2:1
MMR, with periods of $\approx 30$ and $\approx 60$ days. To model the
formation of this system, \cite{LeePeale} performed 3-body simulations
of a central host star and two planets, with additional (dissipative)
forces that reproduced the effects of disc-planet interaction. Only
the outer planet was forced to migrate inward (assuming that the inner
planet has no ambient gas). Capture into resonance occured when the
outer planet crosses the location at which the mean orbital periods
have a ratio of 2. After the resonant capture, however, the
eccentricities of the two planets rise dramatically (except if an
irrealistically high eccentricity damping is applied to the outer
planet).

\section{Migration of a pair of giant planets\,: analytical calculations}

A planet of mass $M_p$, semi-major-axis $a$, and eccentricity $e$ has
an energy $E=-GM_*M_p/2a$ and an angular momentum
$H=M_p\sqrt{GM_*a(1-e)}$, where $M_*$ denote the sum of the masses of
the central star and the planet. Denoting by $\tau_a=-\dot{a}/a$ and
$\tau_e=-\dot{e}/e$, one finds\,:
\begin{eqnarray}
\label{eq:Edot}
\dot{E} & = & E/\tau_a\\
\dot{H} & = & \frac{H}{2}
\left(\frac{2e^2}{1-e^2}\frac{1}{\tau_e}-\frac{1}{\tau_a}\right)\ .
\label{eq:Hdot}
\end{eqnarray}

If two planets are in resonance, the ratio between their energies is
constant. We define\,: $\varepsilon = E_2/E_1 = M_2a_1/M_1a_2$ .
The variation of the energy of the entire system (the pair of
planets), $E=E_1+E_2=E_1(1+\varepsilon)$, is the sum of the energy
variations applied to both planets, such that $\dot{E} =
E_1/\tau_{a_1}+E_2/\tau_{a_2}$.

The same holds for the angular momentum. If the planets are in MMR and
their eccentricities are constant, the ratio between their angular
momenta is also constant\,: $\eta = H_2/H_1$.

With some algebra, one finds a relation between $\tau_{a_1}$,
$\tau_{a_2}$, $\tau_{e_1}$, and $\tau_{e_2}$ for the planets to evolve
in resonance with given eccentricities. It reads\,:
\begin{equation}
\frac{1}{\tau_{e_1}} = \underbrace{\frac{1\!-\!{e_1}^2}{2{e_1}^2} \left( \frac{1}{\tau_{a_2}}\frac{\eta-\varepsilon}{1+\varepsilon} - \frac{1}{\tau_{e_2}}\frac{2{e_2}^2\eta}{1-{e_2}^2}\right) }_{1/\tau_{e_1 \rm max}} + \frac{1}{\tau_{a_1}}\underbrace{\frac{1\!-\!{e_1}^2}{2{e_1}^2}\frac{\varepsilon-\eta}{1+\varepsilon}}_{K_{1}}
\label{eq:tau_e_1}
\end{equation}
The damping rate that should be applied to the inner planet
($1/\tau_{e_1}$) is the sum of two terms. The first one ($1/\tau_{e_1
\rm max}$) is required to balance the action on the outer planet\,; it
is not zero in general. This explains why the eccentricities rised too
much in Lee \& Peale's simulations.

The second term ($K_{1}/\tau_{a_1}$) is proportional to
${\tau_{a_1}}^{\!-1}$, with $K_{1}<0$\,: if the inner planet is pushed
outwards by an inner disk ($\dot{a}_1>0$), an additionnal damping is
required.

\section{Inner disc}

We computed a few hydro-simulations of a giant planet, on a fixed
orbit with various eccentricities. A gaseous disc is present inside
the planetary orbit. We measure the otrque and power of the force of
the disc on the planet. We find that for $e > 0.1$, the disc
damps the eccentricity of the planet. For $e<0.3$, the expected
migration if one releases the planet is directed outwards.

Thus, the inner disc can play a crucial role on the orbital evolution
of a pair of planets in resonance. Note that the evolution of the
inner disc depends crucially on the radius of its inner edge $R_{\rm
inf}$, with respect to $a$, the semi-major-axis of the orbit of the
gap-opening planet \cite{CM07}. If $R_{\rm inf}/a$ is not small with
respect to $1$, the inner disc should disappear promptly. On the other
hand, if $R_{\rm inf}\ll a$, it should not disappear at all.

\section{Application to GJ~876}

\begin{table}
\caption{Parameters of the system GJ~876 as given by
\cite{Laughlin2005} for GJ~876\,b and GJ~876\,c, and by
\texttt{http://exoplanets.eu/} for GJ~876\,d.}
\begin{center}
\begin{tabular}{r|cccc}
\hline
\hline
name & $M.\sin i$ & period &  $a$  & $e$ \\
     & ($M_{\rm Jup}$) & (days) &  (AU) & \\
\hline
b & 1.935 & $60.93\pm 0.03$ & 0.20783 & $0.029 \pm 0.005$\\
c & 0.56 & $30.38\pm 0.03$ & 0.13 & $0.218 \pm 0.002$ \\
d & 0.023 & 1.94 & 0.020807 & 0.
\end{tabular}
\end{center}
\label{tab:gj876}
\end{table}

The two planets GJ~876\,b and GJ~876\,c are in 2:1 MMR (see
table~\ref{tab:gj876}). A third, smaller planet orbits very close to
the star. This planet is not massive enough to open a gap in the gas
disc (see e.g. \cite{CMM06}). Thus it was in type~I migration, and
most likely migrated all the way inwards, until it falls in the empty
cavity between the star and $R_{\rm inf}$. More precisely until the
outermost resonance with the planet (the 2:1) lies in the cavity\,:
the disc and the planet exchange angular momentum through the
resonances. Therefore, we can say that in this system, $R_{\rm inf}$
was located at the 2:1 MMR with GJ~876\,d, that is at $0.033$ AU.

\begin{figure}
\begin{center}
\includegraphics[angle=270,width=0.9\linewidth]{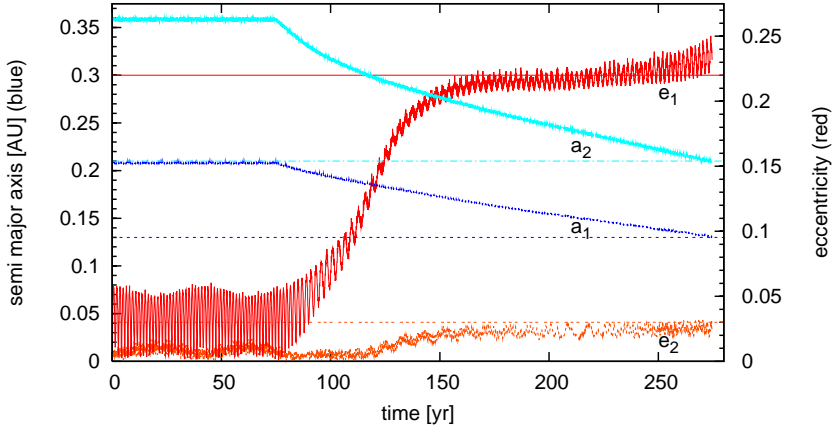}
\caption{Semi-major axis (blue, left $y$-axis) and eccentricity (red,
right $y$-axis) evolution of GJ~876\,b (light colour, $a_1$, $e_1$) and
GJ~876\,c (dark, $a_2$, $e_2$). The horizontal lines correspond to
the observed values, as given by Table~\ref{tab:gj876}.}
\end{center}
\label{fig:GJ876_hydro}
\end{figure}

We use a code in which the whole disc is consistently simulated down
to $R_{\rm inf}$, thanks to the use of 1D grid coupled to the
classical 2D grid \cite{CMM07}. We compute a simulation of the two
most massive planets of GJ~876 in a disc with a viscosity given by an
$\alpha$-prescription with $\alpha=10^{-2}$ and an aspect ratio of
$0.07$. First, the planets are on fixed circular orbits, and shape a
gap in the disc. Then, they are released and they migrate freely. The
results are shown on Figure~\ref{fig:GJ876_hydro}. A resonant capture
in the 2:1 MMR happens at $t=110$\,; then, the eccentricites increase,
but the presence of the inner disc prevents $e$ reaching too high
values. In the end of the simulation, the semi-major-axes and
eccentricities have almost exactly the observed values.

\section{Conclusion}

This work shows that the eccentricity of resonant giant exoplanets is
governed by the presence of a gas disc inside the orbits of the
innermost one. For given eccentricities, resonances and planet masses,
the required action of the inner disc is provided analytically. In the
GJ~876 system, the radius of the inner edge of the disc can be
estimated thanks to the presence of a third planet\,; a numerical
simulation of this system reproduces very well the observed
parameters, which confirms the validity of the model.

In conclusion, the observed moderate eccentricities of the exoplanets
in resonance suggest that in these systems, an inner disc was present
during the migration.

This work has also led to the publication of an article in A\&A
\cite{CSK08}.



\begin{thebibliography}{99}
\bibitem[(Crida \etal, 2008)]{CSK08} Crida, A., S\'andor, Zs, \& Kley, W 2008, A\&A, 483, 325-337
\bibitem[(Crida \& Morbidelli, 2007)]{CM07} Crida, A. \& Morbidelli, A. 2007, MNRAS, 377, 1324-1336
\bibitem[Crida \etal, 2006]{CMM06} Crida, A., Morbidelli, A., \& Masset, F. 2006, Icarus, 181, 587-604
\bibitem[(Crida \etal, 2007)]{CMM07} Crida, A., Morbidelli, M., \& Masset, F. 2007, A\&A, 461, 1173-1183
\bibitem[Laughlin \etal (2005)]{Laughlin2005} Laughlin, G., Butler,~R.~P., Fischer,~D.~A., Marcy,~G.~W., Vogt,~S.~S., Wolf,~A,~S. 2005, ApJ, 622, 1182-1190
\bibitem[Lee \& Peale (2002)]{LeePeale} Lee, M. H. \& Peale, S. J. 2002, ApJ, 567, 596-609
\bibitem[Lin \& Papaloizou, 1986]{LinPapa86} Lin, D. N. C., \& Papaloizou, J. 1986, ApJ, 309, 846-857
\end{thebibliography}
\end{document}